\documentclass[superscriptaddress,twocolumn,prl,aps,showpacs,groupaddress]{revtex4-1}
\usepackage{graphicx}
\usepackage{amsmath}
\usepackage{color}

\newcommand{\ihh}
{Department of Physics, University of Hamburg, 
20355 Hamburg, Germany}

\newcommand{\itap}
{Institut f\"ur Theoretische und Astrophysik,
Christian-Albrechts-Universit\"at zu Kiel,
24098 Kiel, Germany}

\newcommand{\fzj}{Peter Gr\"unberg Institut and Institute for Advanced Simulation,
Forschungszentrum J\"ulich and JARA, Germany}

\newcommand{\jgu}{Johannes Gutenberg-Universit\"at Mainz,
Institute of Physics, Staudingerweg 7, D-55128 Mainz, Germany}

\begin{document}

\title{Competition of Dzyaloshinskii-Moriya and higher-order exchange interactions \\ in Rh/Fe atomic bilayers on Ir(111)}

\author{Niklas Romming}
\affiliation{\ihh}
\author{Henning Pralow}\affiliation{\itap}
\author{Andr\'{e} Kubetzka}\affiliation{\ihh}
\author{Markus Hoffmann}\affiliation{\itap}\affiliation{\fzj}
\author{Stephan von Malottki}\affiliation{\itap}
\author{Sebastian Meyer}\affiliation{\itap}
\author{Bertrand Dup\'{e}}\affiliation{\jgu}
\author{Roland Wiesendanger}\affiliation{\ihh}
\author{Kirsten von Bergmann}\email{Email: kbergman@physnet.uni-hamburg.de}\affiliation{\ihh}
\author{Stefan Heinze}\affiliation{\itap}

\date{\today}

\begin{abstract}
Using spin-polarized scanning tunneling microscopy and density functional theory we demonstrate the occurrence of a novel type of noncollinear spin structure 
in Rh/Fe atomic bilayers on Ir(111). We find that higher-order exchange interactions depend sensitively on the stacking sequence. For fcc-Rh/Fe/Ir(111)
frustrated exchange interactions are dominant and lead to the formation of a spin spiral ground state with a period of about 1.5~nm.
For hcp-Rh/Fe/Ir(111) higher-order exchange interactions favor a double-row wise antiferromagnetic or $\uparrow\uparrow\downarrow\downarrow$ state. 
However, the Dzyaloshinskii-Moriya interaction at the Fe/Ir interface leads to a small angle of about $4^{\circ}$ between adjacent 
magnetic moments resulting in a canted $\uparrow\uparrow\downarrow\downarrow$ ground state.
\end{abstract}

\maketitle

In systems with broken inversion symmetry and strong spin-orbit coupling the Dzyaloshinskii-Moriya interaction (DMI)~\cite{Dzyaloshinskii1957,Moriya1960} plays an essential role for the formation of topologically non-trivial spin structures such as skyrmions \cite{Bogdanov1989,Bogdanov1994,Bogdanov:2001aa,Muhlbauer2009,Yu2010,Seki2012,Fert2013,Nagaosa2013,Wiesendanger2016}.
At transition-metal surfaces and interfaces the DM interaction can induce numerous types
of non-collinear spin structures such as cycloidal spin spirals~\cite{Bode2007,Ferriani2008,Phark2014}, N\'eel type domain 
walls~\cite{Kubetzka2003,Heide2008,Meckler2009,Chen2013a,Emori2013,Ryu2013}, as well as skyrmions and skyrmion 
lattices~\cite{Heinze2011,Romming2013,Romming2015,Chen2015,Jiang2015,Hoffmann2015,Boulle2016,Moreau-Luchaire2016,Woo2016}.

In such systems there is a competition between the DMI favoring a non-collinear spin structure and the Heisenberg exchange, which typically favors collinear alignment of magnetic moments. Depending on their strength and the magnetocrystalline anisotropy energy a spin spiral state forms in zero magnetic field and a transition to skyrmions occurs at finite field. However, higher-order exchange interactions such as the four-spin or biquadratic term can lead to more complex spin structures e.g.~multi-Q states \cite{Kurz2001}, conical spin spirals \cite{PhysRevLett.108.087205} or atomic-scale skyrmion lattices \cite{Heinze2011,Hoffmann2015}. For an Fe monolayer on the Rh(111) surface a so-called up-up-down-down ($\uparrow \uparrow \downarrow \downarrow$) or double-row wise antiferromagnetic state has been predicted \cite{Hardrat:09.1,Al-Zubi:11.2} based on density functional theory (DFT), however, not observed experimentally yet.

Here, we demonstrate that the interplay of the DMI and higher-order exchange can lead to the formation of a novel type of canted
$\uparrow \uparrow \downarrow \downarrow$-state, with small angles between adjacent magnetic moments. We study atomic Rh/Fe bilayers on
the Ir(111) surface which grow pseudomorphically as shown by scanning tunneling microscopy (STM) measurements. 
While Fe grows in fcc stacking both hcp and fcc stacking of Rh are observed.
The ground state spin structure of the film depends on the stacking of the Rh overlayer. In the fcc stacking we observe a spin spiral state in spin-polarized 
(SP-) STM images with a period of 1.5~nm, which is driven by the frustration of exchange interactions as shown from DFT calculations. However, for the hcp stacking of Rh 
an $\uparrow \uparrow \downarrow \downarrow$-state is favorable due to higher-order exchange interactions.
SP-STM shows a stripe pattern with a periodicity close to the four atomic rows spin structure realized in the $\uparrow \uparrow \downarrow \downarrow$-state. 
Using non-spin-polarized STM we find a strong electronic contrast which is explained by the inhomogeneous spin polarization of the Rh overlayer. Spin-polarized 
STM with different tip magnetization directions demonstrates that the spin structure is non-collinear suggesting a small canting of the magnetic moments relative 
to the collinear $\uparrow \uparrow \downarrow \downarrow$-state. We show that this canted spin state is driven by the DMI which is significant at the Fe/Ir interface.

\begin{figure}
\centering
\includegraphics[width=0.9\columnwidth]{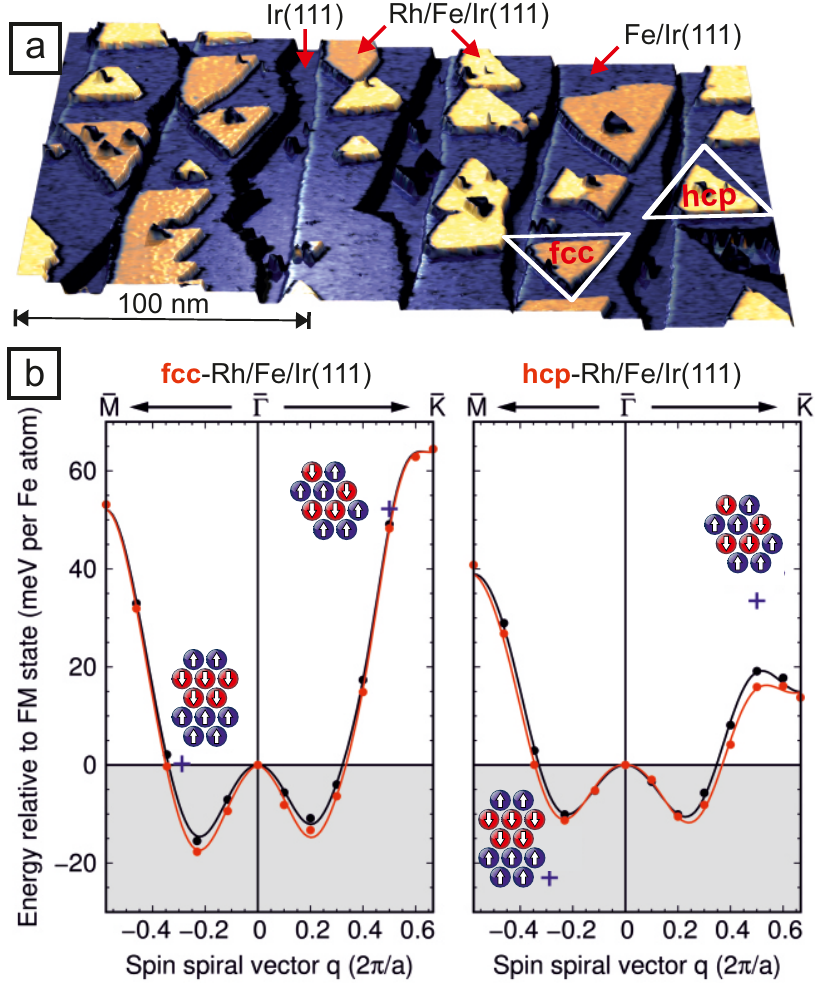}
\caption{\textbf{(a)}
Perspective STM constant-current image of a typical Rh/Fe/Ir(111) sample with approx.~0.8~ML Fe and 0.4~ML Rh, 
colorized with the simultaneously acquired d$I$/d$U$~signal ($U = +0.5$~V, $I = 0.75$~nA, $T = 8$~K).
\textbf{(b)} Calculated energy dispersions $E(\mathbf{q})$ of right-rotating cycloidal homogeneous spin spirals for fcc-Rh (left) and hcp-Rh (right) on Fe/Ir(111) 
without spin-orbit coupling (black symbols) and with spin-orbit coupling (red symbols). The lines denote fits to the Heisenberg model including the 
Dzyaloshinskii-Moriya interaction for the case with spin-orbit coupling (see \cite{SI} for details). 
The energies of the $\uparrow \uparrow \downarrow \downarrow$-states along the two high symmetry directions are marked by blue crosses and
the spin structures are sketched as insets.
}
\label{fig1}
\end{figure}

An STM measurement of a typical Rh/Fe/Ir(111) sample is shown in Fig.\,\ref{fig1}(a), where the color coding refers to the local differential conductance (d$I$/d$U$) (see \cite{SI} for the sample preparation). At this bias voltage, Fe and Ir have similar d$I$/d$U$~signals, but there are two clearly distinguishable contrast levels for the Rh islands. The signal strength correlates with the 
orientation
of the roughly triangular Rh islands, a sign of pseudomorphic growth of Rh with both possible stackings exhibiting slightly different electronic properties. We assign the darker d$I$/d$U$~signal at this bias voltage to fcc-Rh and the brighter one to hcp-Rh (see \cite{SI} for experimental details).

In order to investigate the structural, electronic and magnetic properties of such atomic Rh/Fe bilayers on the Ir(111) surface we have performed DFT calculations using the FLEUR code 
(see \cite{SI} for computational details). We start by discussing the energy dispersion $E({\mathbf q})$ of flat homogeneous spin spirals obtained without taking spin-orbit coupling~(SOC) into account, black data in Fig.\,\ref{fig1}(b). The energy dispersions show that the ferromagnetic (FM) state at $\overline{\Gamma}$ has lower energy than the antiferromagnetic (AFM) state at the Brillouin zone boundary. For both Rh stackings there are deep energy minima on the order of $10-15$~meV/Fe atom for spin spirals with periods of $\lambda \approx 1.2$~nm. The origin of these spin spiral minima is frustration of exchange interactions, where the nearest-neighbor exchange interaction favors FM alignment, but 2nd or 3rd nearest neighbor exchange interactions are AFM 
(for exchange constants see \cite{SI}).

Including spin-orbit coupling (SOC), 
see red data in Fig.\,\ref{fig1}(b), leads to a preference of right rotating cycloidal spin spirals due to DMI for both types of stacking of the Rh overlayer. However, the energy differences are relatively small compared to the depths of the spin spiral energy minima neglecting SOC. 

Because a significant role of higher-order exchange interactions has been reported for the Fe/Rh~\cite{Hardrat:09.1,Al-Zubi:11.2} and Fe/Ir~\cite{Heinze2011} interfaces, we have also considered 
collinear $\uparrow \uparrow \downarrow \downarrow$-states along the high symmetry directions. These states are formed by the superposition of spin spirals and
should be energetically degenerate with them within the Heisenberg model. Energy differences obtained within DFT indicate higher-order exchange contributions.
For fcc-Rh we find that both $\uparrow \uparrow \downarrow \downarrow$-states have a higher energy compared to the respective spin spirals, and the magnetic ground state remains a spin spiral along $\overline{\Gamma \mathrm{M}}$ [Fig.\,\ref{fig1}(b)].
For hcp-Rh we find that the $\uparrow \uparrow \downarrow \downarrow$-state along the $\overline{\Gamma \mathrm{K}}$ direction is 
about 34 meV/Fe-atom higher 
than the FM state, however, the $\uparrow \uparrow \downarrow \downarrow$-state along the $\overline{\Gamma \mathrm{M}}$ direction is by about 12~meV/Fe-atom lower in energy than the lowest spin spiral state [Fig.\,\ref{fig1}(b)]. 

\begin{figure}
\centering
\includegraphics[width=0.9\columnwidth]{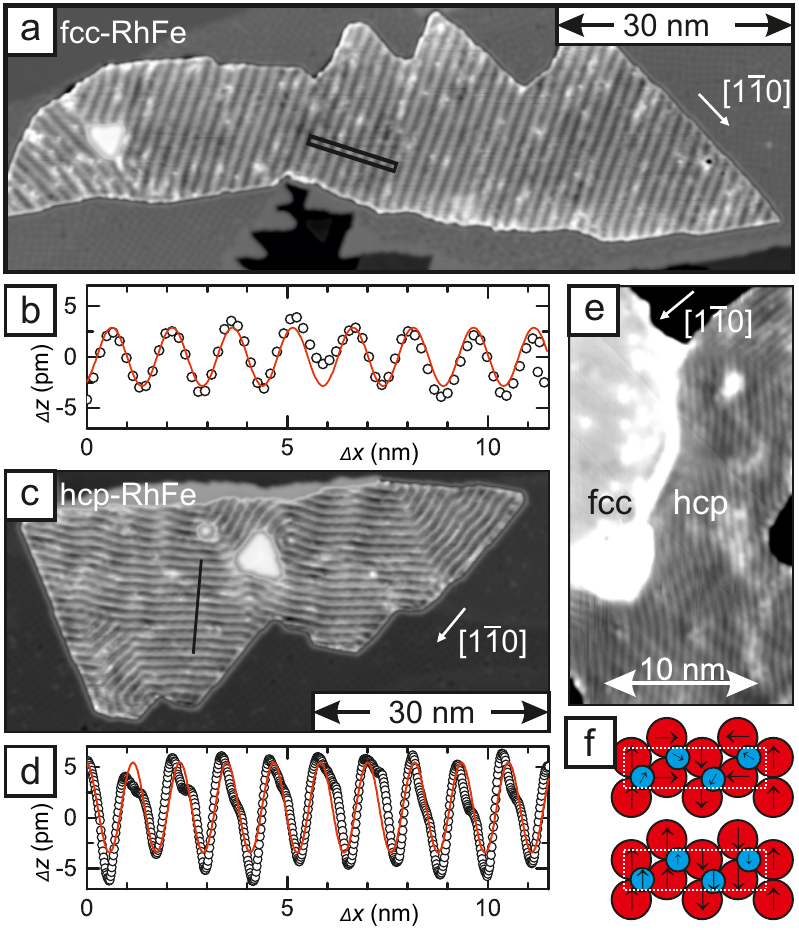}
\caption{\textbf{(a)},\textbf{(c)}~SP-STM constant-current images of an fcc-Rh and an hcp-Rh island on Fe/Ir(111), respectively, measured with a magnetic Cr~tip, sensitive to the out-of-plane magnetization component of the sample (contrast level adjusted locally on the Rh/Fe to $\pm 20$~pm; $T = 8$~K, $U = +30$ and $-30$~mV, $I = 1.0$ and $1.5$~nA, respectively). \textbf{(b)},\textbf{(d)}~Line profiles along the rectangles in (a,c); solid lines are fits with cosine-functions. \textbf{(e)}~Constant-current STM image of a Rh island with fcc and hcp stacking on fcc-Fe/Ir(111) measured with a non-spin-polarized tip (contrast $\pm 15$~pm; $T = 8$~K, $U = +15$~mV, $I = 6$~nA). \textbf{(f)}~Top-view sketches of a homogeneous 4 atom period spin spiral and the $\uparrow \uparrow \downarrow \downarrow$-state; Fe atoms in red and Rh atoms in blue, the magnetic unit cell is indicated by the dotted box; for simplicity in-plane magnetic states are sketched.}\label{fig2}
\end{figure}

When a magnetic tip is used in STM the tunnel magnetoresistance (TMR) effect occurs, which leads to a SP contribution to the tunnel current in addition to the structural and electronic part~\cite{Bode2003a,Wiesendanger2009}. Figure~\ref{fig2}(a) shows such an SP-STM measurement of an fcc-Rh/Fe island exhibiting a cosine-like magnetic modulation with a period $\lambda_{\mathrm{fcc}}\approx 1.5$~nm (see Fig.~\ref{fig2}(b)) and propagation along $\left< 11\bar{2} \right>$ directions. We conclude that fcc-Rh/Fe exhibits a spin spiral ground state with a continuous rotation of adjacent magnetic moments (see \cite{SI} for further measurements), in agreement with the spin spiral energy minimum found along the $\overline{\Gamma \mathrm{M}}$ direction from DFT (cf.~Fig.~\ref{fig2}(a)) 
\cite{spindynamics}.

The SP-STM image in Fig.~\ref{fig2}(c) shows an hcp-Rh/Fe island and a similar stripe pattern with a slightly smaller period \mbox{($\lambda_{\mathrm{hcp}}\approx 1.1$~nm)} is visible. The magnetic structure differs from the one in fcc-Rh/Fe islands in subtle aspects: the propagation direction seems to be more flexible, is not strictly along $\left< 11\bar{2} \right>$ directions but instead varies locally; the shape of the periodic signal significantly differs from a cosine, compare profile in Fig.~\ref{fig2}(d). 
This demonstrates that the magnetic ground state of the hcp-Rh/Fe island is different to that of fcc-Rh/Fe.

Because in STM several different magnetoresistive (MR) 
effects can contribute to the measurement signal~\cite{Bode2002,Hanneken:15.1}, in Fig.\,\ref{fig2}(e) we use a non-spin-polarized tip to separate purely electronic contributions from signal variations due to the TMR. 
We find that for fcc and hcp-Rh the 
TMR signals with periods of about $1.5$~nm and $1.1$~nm, respectively, vanish, meaning that they originate from TMR. On fcc-Rh on Fe/Ir(111) no remaining modulation of the signal is observed in the bias voltage regime $\pm1$~V. In contrast, the non-spin-polarized signal observed on hcp-Rh is rather strong, i.e.\ on the order of a few pm, with half of the magnetic period, see Fig.\,\ref{fig2}(e), and can be observed in a bias voltage regime of around $\pm 0.2$~V. 

Regarding the purely electronic contributions to magnetoresistance, the generic differences between a spin spiral and an $\uparrow \uparrow \downarrow \downarrow$-state, as predicted by DFT for our atomic Rh/Fe bilayer system, can be understood based on the sketches in Fig.\,\ref{fig2}(f). The tunnel anisotropic magnetoresistance (TAMR)~\cite{Bode2002} due to spin-orbit coupling can occur for spin spirals (in-plane spins are inequivalent to out-of-plane spins) and would manifest as a modulation with half of the magnetic period, but it is absent in a collinear magnetic state as the $\uparrow \uparrow \downarrow \downarrow$-state. Regarding the non-collinear magnetoresistance~\cite{Hanneken:15.1}, the four-atom magnetic period is a special case where the contribution to the total 
magnetoresistance does not vary locally. 

\begin{figure}
\centering
\includegraphics[width=0.9\columnwidth]{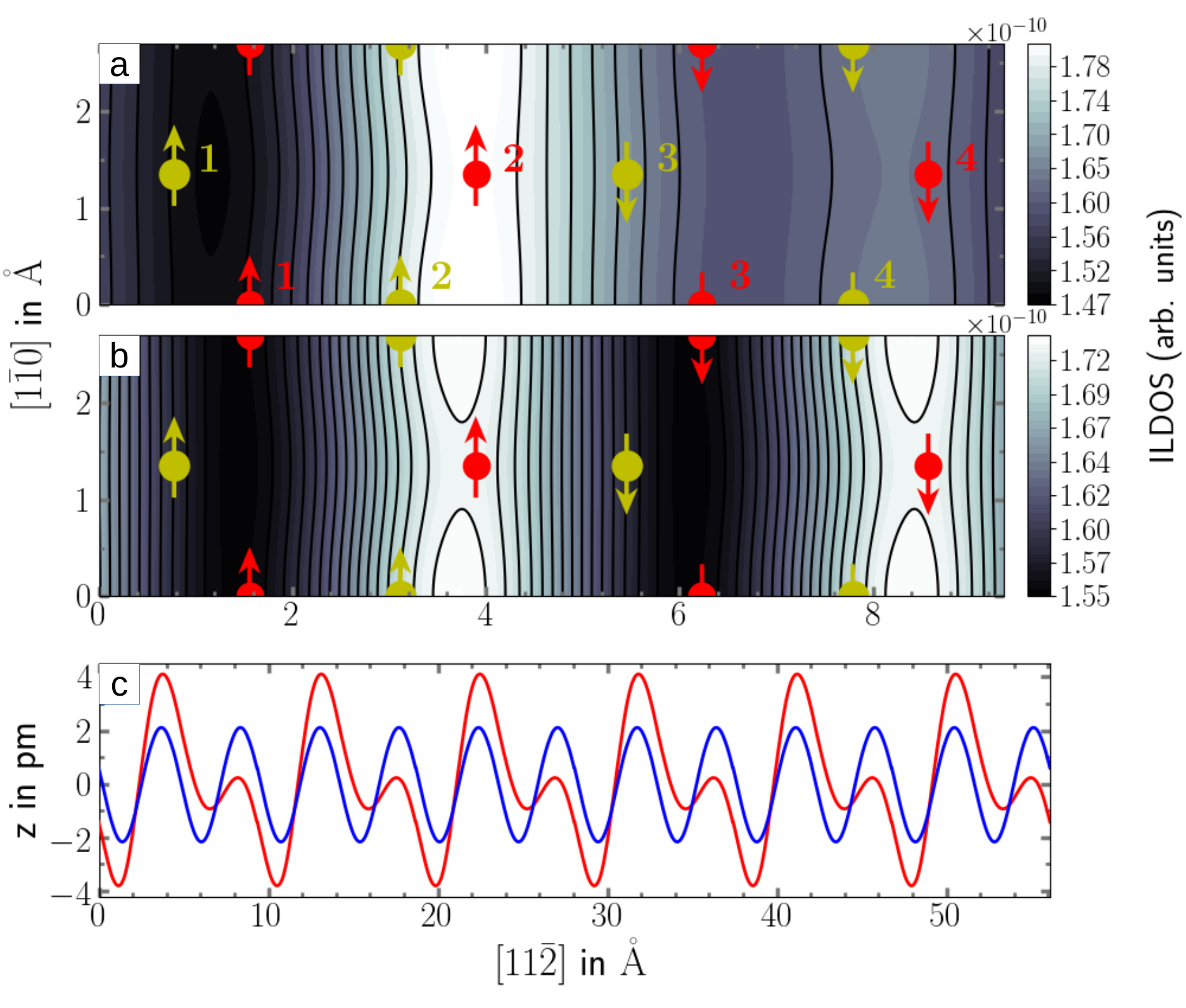}
\caption{STM images simulated based on the DFT calculations for the $\uparrow \uparrow \downarrow \downarrow$ state along $\overline{\Gamma \mathrm{M}}$ of hcp-Rh/Fe/Ir(111).
\textbf{(a)} STM image at a tip-sample distance of $z=6.7$~{\AA} and for a tip spin polarization of $0.5$. The red and yellow filled circles represent the Fe and Rh atoms, respectively, and the arrows show the direction of the magnetic moments in the $\uparrow \uparrow \downarrow \downarrow$ state. Note that the moments have been drawn in the film plane for illustration although the easy axis is perpendicular to the film. \textbf{(b)} simulated STM image for a non-spin-polarized STM tip. \textbf{(c)} Line scans along the $[11\overline{2}]$ direction for the simulated STM images in \textbf{(a)} (red line) and in (b) (blue line). States between the Fermi energy $E_F$ and 0.1~eV below $E_F$ have been taken into account for the simulated STM images.}
\label{fig:STM_simulation_DFT}
\end{figure}

We can calculate STM images for the $\uparrow \uparrow \downarrow \downarrow$-state of hcp-Rh/Fe/Ir(111) directly from DFT. Within the Tersoff-Hamann 
model \cite{TH-model} and its generalization to the spin-polarized case \cite{PhysRevLett.86.4132} the STM image corresponds to the (spin-resolved) local density 
of states (LDOS) a few {\AA} above the surface. The simulated SP-STM image [Fig.~\ref{fig:STM_simulation_DFT}(a)] assuming a small negative bias voltage
of 0.1~V shows a stripe pattern with the magnetic period corresponding to the $\uparrow \uparrow \downarrow \downarrow$-state. The scan lines display an
asymmetric shape as seen in Fig.~\ref{fig:STM_simulation_DFT}(c). If we assume a vanishing spin-polarization of the tip in our calculations we obtain the 
STM image in Fig.~\ref{fig:STM_simulation_DFT}(b). We find a stripe pattern with half the magnetic period as in the experiment. As shown by the scan lines 
given in Fig.~\ref{fig:STM_simulation_DFT}(c), the corrugation amplitude amounts to a few pm. 

This electronic effect has been predicted before based on DFT 
for the $\uparrow \uparrow \downarrow \downarrow$-state in an Fe monolayer on Rh(111)~\cite{Al-Zubi:11.2}. For Rh/Fe/Ir(111) the effect is much enhanced
as it originates from the interface with the Rh layer which is at the surface in our case. 
The $\uparrow \uparrow \downarrow \downarrow$-state can be viewed as the extreme limit of an inhomogeneous spin spiral, and 
the Rh atoms become inequivalent and have different magnetic moments depending on whether all three neighboring Fe atoms are parallel 
($m_{\rm Rh1}^{\rm hcp}=\pm 0.43\mu_{\rm B}$) or two are antiparallel and only one is parallel ($m_{\rm Rh2}^{\rm hcp}=\pm 0.08 \mu_{\rm B}$)
(see sketch in Fig.~\ref{fig2}f).

\begin{figure}
\centering
\includegraphics[width=0.9\columnwidth]{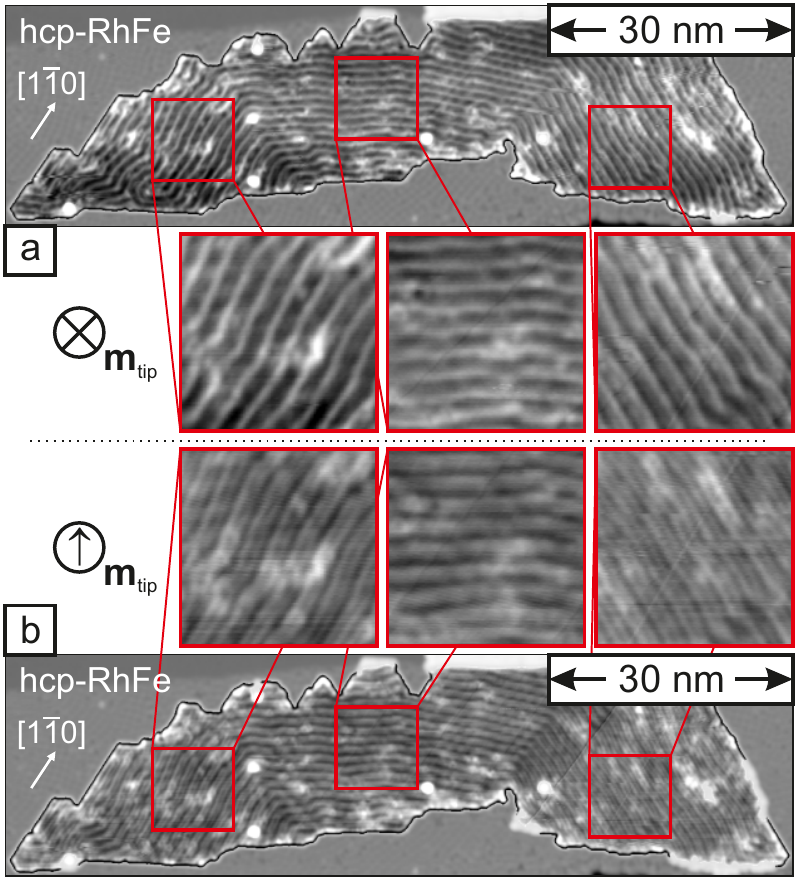}
\caption{\textbf{(a)},\textbf{(b)}~Constant-current SP-STM images of a hcp-Rh island on Fe/Ir(111) measured with out-of-plane sensitive and in-plane sensitive magnetic Fe-coated W tip, respectively; the tip magnetization $\mathbf{m}_{tip}$ aligns in the applied out-of-plane field of $B = -2$~T (a), but is in the sample surface plane without applied field (b); its direction as indicated is derived from comparison of the relative magnetic contrast amplitudes of the three symmetry-equivalent rotational domains, see enlarged images (all contrast levels adjusted locally to $\pm 15$~pm) ($T = 8$~K, $U = +30$~mV, $I = 3$~nA).}\label{fig3}
\end{figure}

To investigate experimentally whether the hcp-Rh/Fe exhibits a strictly collinear magnetic ground state we image the out-of-plane and the in-plane magnetization components of the same island, see Fig.\,\ref{fig3}(a) and (b), respectively. This is done by using an Fe-coated W tip that aligns its magnetization 
in out-of-plane magnetic fields and thus detects out-of-plane sample magnetization components, whereas it has a magnetization in the surface plane without external magnetic field, thus being sensitive to the in-plane magnetization components of the sample. Figure~\ref{fig3}(a) demonstrates that in measurements with an out-of-plane sensitive tip all three rotational domains appear the same; the pattern consists of slim bright lines spaced with the magnetic periodicity. Such a pattern is observed when the magnetic and the electronic MR contributions are of similar magnitude and in phase, i.e.\ the magnetic maxima/minima coincide with the maxima of the electronic contribution and thus add-up/cancel~\cite{Bergmann2012}.

The SP-STM image with in-plane magnetized tip in Fig.\,\ref{fig3}(b) shows a qualitatively different pattern in the central rotational domain compared to the two other rotational domains. This immediately means that there are also magnetic in-plane components in the sample; given that the spin texture is of cycloidal nature due to the DMI, we can derive a tip magnetization axis as indicated, leading to a large magnetic contribution to the signal for the central domain. The magnetic signal is small in the other two domains and there the electronic contribution with half the magnetic period dominates the image. From these measurements we conclude that we have both out-of-plane as well as in-plane magnetization components in this sample and can thus rule out a strictly collinear 
$\uparrow \uparrow \downarrow \downarrow$ state. 
However, because of the large electronic effect observed experimentally, which we attribute to the polarization variation of the Rh atoms, we propose that the hcp-Rh/Fe/Ir(111) realizes a magnetic state in between the two extreme cases of homogeneous spin spiral and $\uparrow \uparrow \downarrow \downarrow$ (see Fig.\,\ref{fig2}(f)), i.e.\ an inhomogeneous spin spiral or a canted $\uparrow \uparrow \downarrow \downarrow$-state. Because the periodic modulation of the LDOS as manifested in the electronic contrast also changes the spin-polarization of the Rh atoms (see Fig.~S6 \cite{SI}), we cannot quantitatively compare the measured magnetic amplitudes to extract the canting angle.

\begin{figure}
\centering
\includegraphics[width=0.9\columnwidth]{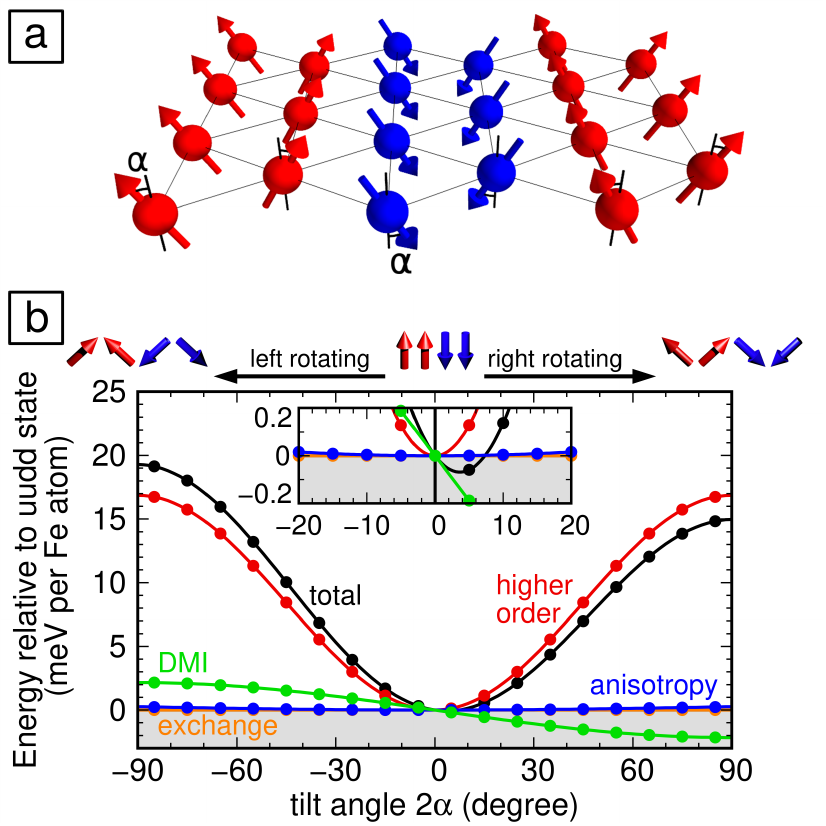}
\caption{\textbf{(a)}~Sketch of the canted $\uparrow \uparrow \downarrow \downarrow$ state defined by the angle $\alpha$. Only the magnetic moments in the Fe layer are shown. Blue and red denotes an 
up- or downwards out-of-plane magnetization component. 
\textbf{(b)}~Energy as a function of $\alpha$ resolved by the contributions from different magnetic interactions. The filled circles are obtained numerically using the DFT interaction
parameters for the atomistic spin model and the lines are from the analytical forms of the energy contributions (see text). Note that positive/negative values of $\alpha$ indicate an anti-clockwise/clockwise rotation of the magnetic moments.}\label{fig4}
\end{figure}

The DFT calculations of Fig.\,\ref{fig1}(b) have considered homogeneous spin spirals and collinear $\uparrow \uparrow \downarrow \downarrow$-states. 
To include inhomogeneous spin spirals, as found experimentally, and study which energy contributions could lead to such a state, we introduce a canting angle $\alpha$ relative 
to the easy magnetization axis, see Fig.\,\ref{fig4}(a). This allows to continuously transform the $\uparrow \uparrow \downarrow \downarrow$-state ($\alpha=0^{\circ}$) to the $90^{\circ}$ clockwise spin spiral ($\alpha=45^{\circ}$). Negative values of $\alpha$ denote an anti-clockwise spin canting.
It can easily be shown that the exchange energy contribution does not depend on $\alpha$. The magnetocrystalline anisotropy energy stabilizes the collinear state 
and varies as $E_{\rm MAE} \propto -\cos^2{\alpha}$, however, the MAE is small with 0.53~meV/Fe-atom.  
The DMI favors a canting and decreases as $E_{\rm DMI} \propto -\sin{2 \alpha}$ for clockwise spin canting (Fig.\,\ref{fig4}(b)). Note that the strength of the DMI and the MAE are obtained from DFT.
The energy difference of about 17~meV/Fe-atom between the $\uparrow \uparrow \downarrow \downarrow$ and the $90^{\circ}$ spin spiral state can only be due to higher-order exchange contributions
since spin-orbit 
coupling has been turned off in the DFT calculation. This leads to a rise with $(1-\cos^2{\alpha})$ if we assume only nearest-neighbor four-spin and biquadratic interaction 
(see Eqs.~(3) and (4) in \cite{SI}).
The competition of DMI and these higher-order contributions leads to an energy minimum at a canting angle of $2\alpha \approx 4 ^{\circ}$ between adjacent spins which are parallel in the 
collinear $\uparrow \uparrow \downarrow \downarrow$-state. 

To check the validity of the spin model we have performed self-consistent non-collinear DFT calculations including spin-orbit coupling in the four atom per layer super cell of the 
$\uparrow \uparrow \downarrow \downarrow$ state allowing the spins to relax to find the energetically most favorable state \cite{Kurz2004}. To make these calculations computationally
feasible, we have considered a freestanding Rh/Fe/Ir trilayer which is very similar to the Rh/Fe/Ir(111) film system in terms of its magnetic properties (see \cite{SI} for details). 
We find a canted $\uparrow \uparrow \downarrow \downarrow$ state with
$2\alpha \approx 7 ^{\circ}$ energetically more favorable by 0.03~meV/Fe-atom than the collinear $\uparrow \uparrow \downarrow \downarrow$-state. Two of the Rh magnetic moments point almost 
perpendicular to the surface while the other two are at angles of about $13^\circ$ with respect to the surface normal (see Fig.~S4 \cite{SI}). Thus the in-plane components are enhanced at the Rh surface layer which explains the relatively strong in-plane contrast observed in SP-STM measurements (cf.~Fig.~S7 \cite{SI}).


In conclusion, we have shown that higher-order exchange interactions can play a decisive role in transition-metal trilayers and may compete
with interfacial DM interactions. Our work demonstrates that higher-order exchange needs to be taken into account in the search for
novel transition-metal interfaces potentially promising for complex non-collinear spin structures such as skyrmions. 

\begin{acknowledgments}
This work has received financial support by the European Union via the Horizon 2020 research and innovation programme under grant agreement No.~665095 (FET-Open project MagicSky), and by the Deutsche Forschungsgemeinschaft via SFB668-A8. We gratefully acknowledge computing time at the supercomputer of the North-German Supercomputing Alliance (HLRN).
\end{acknowledgments}


\begin{thebibliography}{43}
\expandafter\ifx\csname natexlab\endcsname\relax\def\natexlab#1{#1}\fi
\expandafter\ifx\csname bibnamefont\endcsname\relax
  \def\bibnamefont#1{#1}\fi
\expandafter\ifx\csname bibfnamefont\endcsname\relax
  \def\bibfnamefont#1{#1}\fi
\expandafter\ifx\csname citenamefont\endcsname\relax
  \def\citenamefont#1{#1}\fi
\expandafter\ifx\csname url\endcsname\relax
  \def\url#1{\texttt{#1}}\fi
\expandafter\ifx\csname urlprefix\endcsname\relax\def\urlprefix{URL }\fi
\providecommand{\bibinfo}[2]{#2}
\providecommand{\eprint}[2][]{\url{#2}}

\bibitem[{\citenamefont{Dzyaloshinskii}(1957)}]{Dzyaloshinskii1957}
\bibinfo{author}{\bibfnamefont{I.~E.} \bibnamefont{Dzyaloshinskii}},
  \bibinfo{journal}{Sov. Phys. JETP} \textbf{\bibinfo{volume}{5}},
  \bibinfo{pages}{1259} (\bibinfo{year}{1957}).

\bibitem[{\citenamefont{Moriya}(1960)}]{Moriya1960}
\bibinfo{author}{\bibfnamefont{T.}~\bibnamefont{Moriya}},
  \bibinfo{journal}{Phys. Rev.} \textbf{\bibinfo{volume}{120}},
  \bibinfo{pages}{91} (\bibinfo{year}{1960}).

\bibitem[{\citenamefont{Bogdanov and Yablonskii}(1989)}]{Bogdanov1989}
\bibinfo{author}{\bibfnamefont{A.}~\bibnamefont{Bogdanov}} \bibnamefont{and}
  \bibinfo{author}{\bibfnamefont{D.~A.} \bibnamefont{Yablonskii}},
  \bibinfo{journal}{Sov. Phys. JETP} \textbf{\bibinfo{volume}{68}},
  \bibinfo{pages}{101} (\bibinfo{year}{1989}).

\bibitem[{\citenamefont{Bogdanov and Hubert}(1994)}]{Bogdanov1994}
\bibinfo{author}{\bibfnamefont{A.}~\bibnamefont{Bogdanov}} \bibnamefont{and}
  \bibinfo{author}{\bibfnamefont{A.}~\bibnamefont{Hubert}},
  \bibinfo{journal}{J. Mag. Mag. Mat.} \textbf{\bibinfo{volume}{138}},
  \bibinfo{pages}{255} (\bibinfo{year}{1994}).

\bibitem[{\citenamefont{Bogdanov and R{\"o}{\ss}ler}(2001)}]{Bogdanov:2001aa}
\bibinfo{author}{\bibfnamefont{A.~N.} \bibnamefont{Bogdanov}} \bibnamefont{and}
  \bibinfo{author}{\bibfnamefont{U.~K.} \bibnamefont{R{\"o}{\ss}ler}},
  \bibinfo{journal}{Phys. Rev. Lett.} \textbf{\bibinfo{volume}{87}},
  \bibinfo{pages}{037203} (\bibinfo{year}{2001}).

\bibitem[{\citenamefont{M\"uhlbauer et~al.}(2009)\citenamefont{M\"uhlbauer,
  Binz, Jonietz, Pfleiderer, Rosch, Neubauer, Georgii, and
  B\"oni}}]{Muhlbauer2009}
\bibinfo{author}{\bibfnamefont{S.}~\bibnamefont{M\"uhlbauer}},
  \bibinfo{author}{\bibfnamefont{B.}~\bibnamefont{Binz}},
  \bibinfo{author}{\bibfnamefont{F.}~\bibnamefont{Jonietz}},
  \bibinfo{author}{\bibfnamefont{C.}~\bibnamefont{Pfleiderer}},
  \bibinfo{author}{\bibfnamefont{A.}~\bibnamefont{Rosch}},
  \bibinfo{author}{\bibfnamefont{A.}~\bibnamefont{Neubauer}},
  \bibinfo{author}{\bibfnamefont{R.}~\bibnamefont{Georgii}}, \bibnamefont{and}
  \bibinfo{author}{\bibfnamefont{P.}~\bibnamefont{B\"oni}},
  \bibinfo{journal}{Science} \textbf{\bibinfo{volume}{323}},
  \bibinfo{pages}{915} (\bibinfo{year}{2009}).

\bibitem[{\citenamefont{Yu et~al.}(2010)\citenamefont{Yu, Onose, Kanazawa,
  Park, Han, Matsui, Nagaosa, and Tokura}}]{Yu2010}
\bibinfo{author}{\bibfnamefont{X.~Z.} \bibnamefont{Yu}},
  \bibinfo{author}{\bibfnamefont{Y.}~\bibnamefont{Onose}},
  \bibinfo{author}{\bibfnamefont{N.}~\bibnamefont{Kanazawa}},
  \bibinfo{author}{\bibfnamefont{J.~H.} \bibnamefont{Park}},
  \bibinfo{author}{\bibfnamefont{J.~H.} \bibnamefont{Han}},
  \bibinfo{author}{\bibfnamefont{Y.}~\bibnamefont{Matsui}},
  \bibinfo{author}{\bibfnamefont{N.}~\bibnamefont{Nagaosa}}, \bibnamefont{and}
  \bibinfo{author}{\bibfnamefont{Y.}~\bibnamefont{Tokura}},
  \bibinfo{journal}{Nature} \textbf{\bibinfo{volume}{465}},
  \bibinfo{pages}{901} (\bibinfo{year}{2010}).

\bibitem[{\citenamefont{Seki et~al.}(2012)\citenamefont{Seki, Yu, Ishiwata, and
  Tokura}}]{Seki2012}
\bibinfo{author}{\bibfnamefont{S.}~\bibnamefont{Seki}},
  \bibinfo{author}{\bibfnamefont{X.~Z.} \bibnamefont{Yu}},
  \bibinfo{author}{\bibfnamefont{S.}~\bibnamefont{Ishiwata}}, \bibnamefont{and}
  \bibinfo{author}{\bibfnamefont{Y.}~\bibnamefont{Tokura}},
  \bibinfo{journal}{Science} \textbf{\bibinfo{volume}{336}},
  \bibinfo{pages}{198} (\bibinfo{year}{2012}).

\bibitem[{\citenamefont{Fert et~al.}(2013)\citenamefont{Fert, Cros, and
  Sampaio}}]{Fert2013}
\bibinfo{author}{\bibfnamefont{A.}~\bibnamefont{Fert}},
  \bibinfo{author}{\bibfnamefont{V.}~\bibnamefont{Cros}}, \bibnamefont{and}
  \bibinfo{author}{\bibfnamefont{J.}~\bibnamefont{Sampaio}},
  \bibinfo{journal}{Nat. Nanotechnol.} \textbf{\bibinfo{volume}{8}},
  \bibinfo{pages}{152} (\bibinfo{year}{2013}).

\bibitem[{\citenamefont{Nagaosa and Tokura}(2013)}]{Nagaosa2013}
\bibinfo{author}{\bibfnamefont{N.}~\bibnamefont{Nagaosa}} \bibnamefont{and}
  \bibinfo{author}{\bibfnamefont{Y.}~\bibnamefont{Tokura}},
  \bibinfo{journal}{Nat. Nanotechnol.} \textbf{\bibinfo{volume}{8}},
  \bibinfo{pages}{899} (\bibinfo{year}{2013}).

\bibitem[{\citenamefont{Wiesendanger}(2016)}]{Wiesendanger2016}
\bibinfo{author}{\bibfnamefont{R.}~\bibnamefont{Wiesendanger}},
  \bibinfo{journal}{Nat. Rev. Mat.} \textbf{\bibinfo{volume}{1}},
  \bibinfo{pages}{16044} (\bibinfo{year}{2016}).

\bibitem[{\citenamefont{Bode et~al.}(2007)\citenamefont{Bode, Heide, von
  Bergmann, Ferriani, Heinze, Bihlmayer, Kubetzka, Pietzsch, Bl{\"{u}}gel, and
  Wiesendanger}}]{Bode2007}
\bibinfo{author}{\bibfnamefont{M.}~\bibnamefont{Bode}},
  \bibinfo{author}{\bibfnamefont{M.}~\bibnamefont{Heide}},
  \bibinfo{author}{\bibfnamefont{K.}~\bibnamefont{von Bergmann}},
  \bibinfo{author}{\bibfnamefont{P.}~\bibnamefont{Ferriani}},
  \bibinfo{author}{\bibfnamefont{S.}~\bibnamefont{Heinze}},
  \bibinfo{author}{\bibfnamefont{G.}~\bibnamefont{Bihlmayer}},
  \bibinfo{author}{\bibfnamefont{a.}~\bibnamefont{Kubetzka}},
  \bibinfo{author}{\bibfnamefont{O.}~\bibnamefont{Pietzsch}},
  \bibinfo{author}{\bibfnamefont{S.}~\bibnamefont{Bl{\"{u}}gel}},
  \bibnamefont{and}
  \bibinfo{author}{\bibfnamefont{R.}~\bibnamefont{Wiesendanger}},
  \bibinfo{journal}{Nature} \textbf{\bibinfo{volume}{447}},
  \bibinfo{pages}{190} (\bibinfo{year}{2007}).

\bibitem[{\citenamefont{Ferriani et~al.}(2008)\citenamefont{Ferriani, von
  Bergmann, Vedmedenko, Heinze, Bode, Heide, Bihlmayer, Bluegel, and
  Wiesendanger}}]{Ferriani2008}
\bibinfo{author}{\bibfnamefont{P.}~\bibnamefont{Ferriani}},
  \bibinfo{author}{\bibfnamefont{K.}~\bibnamefont{von Bergmann}},
  \bibinfo{author}{\bibfnamefont{E.~Y.} \bibnamefont{Vedmedenko}},
  \bibinfo{author}{\bibfnamefont{S.}~\bibnamefont{Heinze}},
  \bibinfo{author}{\bibfnamefont{M.}~\bibnamefont{Bode}},
  \bibinfo{author}{\bibfnamefont{M.}~\bibnamefont{Heide}},
  \bibinfo{author}{\bibfnamefont{G.}~\bibnamefont{Bihlmayer}},
  \bibinfo{author}{\bibfnamefont{S.}~\bibnamefont{Bluegel}}, \bibnamefont{and}
  \bibinfo{author}{\bibfnamefont{R.}~\bibnamefont{Wiesendanger}},
  \bibinfo{journal}{Phys. Rev. Lett.} \textbf{\bibinfo{volume}{101}},
  \bibinfo{pages}{027201} (\bibinfo{year}{2008}).

\bibitem[{\citenamefont{Phark et~al.}(2014)\citenamefont{Phark, Fischer,
  Corbetta, Sander, Nakamura, and Kirschner}}]{Phark2014}
\bibinfo{author}{\bibfnamefont{S.~H.} \bibnamefont{Phark}},
  \bibinfo{author}{\bibfnamefont{J.~A.} \bibnamefont{Fischer}},
  \bibinfo{author}{\bibfnamefont{M.}~\bibnamefont{Corbetta}},
  \bibinfo{author}{\bibfnamefont{D.}~\bibnamefont{Sander}},
  \bibinfo{author}{\bibfnamefont{K.}~\bibnamefont{Nakamura}}, \bibnamefont{and}
  \bibinfo{author}{\bibfnamefont{J.}~\bibnamefont{Kirschner}},
  \bibinfo{journal}{Nat. Commun.} \textbf{\bibinfo{volume}{5}},
  \bibinfo{pages}{5183} (\bibinfo{year}{2014}).

\bibitem[{\citenamefont{Kubetzka et~al.}(2003)\citenamefont{Kubetzka, Pietzsch,
  Bode, and Wiesendanger}}]{Kubetzka2003}
\bibinfo{author}{\bibfnamefont{A.}~\bibnamefont{Kubetzka}},
  \bibinfo{author}{\bibfnamefont{O.}~\bibnamefont{Pietzsch}},
  \bibinfo{author}{\bibfnamefont{M.}~\bibnamefont{Bode}}, \bibnamefont{and}
  \bibinfo{author}{\bibfnamefont{R.}~\bibnamefont{Wiesendanger}},
  \bibinfo{journal}{Phys. Rev. B} \textbf{\bibinfo{volume}{67}},
  \bibinfo{pages}{020401} (\bibinfo{year}{2003}).

\bibitem[{\citenamefont{Heide et~al.}(2008)\citenamefont{Heide, Bihlmayer, and
  Bl{\"{u}}gel}}]{Heide2008}
\bibinfo{author}{\bibfnamefont{M.}~\bibnamefont{Heide}},
  \bibinfo{author}{\bibfnamefont{G.}~\bibnamefont{Bihlmayer}},
  \bibnamefont{and}
  \bibinfo{author}{\bibfnamefont{S.}~\bibnamefont{Bl{\"{u}}gel}},
  \bibinfo{journal}{Phys. Rev. B} \textbf{\bibinfo{volume}{78}},
  \bibinfo{pages}{140403} (\bibinfo{year}{2008}).

\bibitem[{\citenamefont{Meckler et~al.}(2009)\citenamefont{Meckler, Mikuszeit,
  Pressler, Vedmedenko, Pietzsch, and Wiesendanger}}]{Meckler2009}
\bibinfo{author}{\bibfnamefont{S.}~\bibnamefont{Meckler}},
  \bibinfo{author}{\bibfnamefont{N.}~\bibnamefont{Mikuszeit}},
  \bibinfo{author}{\bibfnamefont{A.}~\bibnamefont{Pressler}},
  \bibinfo{author}{\bibfnamefont{E.~Y.} \bibnamefont{Vedmedenko}},
  \bibinfo{author}{\bibfnamefont{O.}~\bibnamefont{Pietzsch}}, \bibnamefont{and}
  \bibinfo{author}{\bibfnamefont{R.}~\bibnamefont{Wiesendanger}},
  \bibinfo{journal}{Phys. Rev. Lett.} \textbf{\bibinfo{volume}{103}},
  \bibinfo{pages}{157201} (\bibinfo{year}{2009}).

\bibitem[{\citenamefont{Chen et~al.}(2013)\citenamefont{Chen, Zhu, Quesada, Li,
  N'Diaye, Huo, Ma, Chen, Kwon, Won et~al.}}]{Chen2013a}
\bibinfo{author}{\bibfnamefont{G.}~\bibnamefont{Chen}},
  \bibinfo{author}{\bibfnamefont{J.}~\bibnamefont{Zhu}},
  \bibinfo{author}{\bibfnamefont{A.}~\bibnamefont{Quesada}},
  \bibinfo{author}{\bibfnamefont{J.}~\bibnamefont{Li}},
  \bibinfo{author}{\bibfnamefont{A.~T.} \bibnamefont{N'Diaye}},
  \bibinfo{author}{\bibfnamefont{Y.}~\bibnamefont{Huo}},
  \bibinfo{author}{\bibfnamefont{T.~P.} \bibnamefont{Ma}},
  \bibinfo{author}{\bibfnamefont{Y.}~\bibnamefont{Chen}},
  \bibinfo{author}{\bibfnamefont{H.~Y.} \bibnamefont{Kwon}},
  \bibinfo{author}{\bibfnamefont{C.}~\bibnamefont{Won}}, \bibnamefont{et~al.},
  \bibinfo{journal}{Phys. Rev. Lett.} \textbf{\bibinfo{volume}{110}},
  \bibinfo{pages}{177204} (\bibinfo{year}{2013}).

\bibitem[{\citenamefont{Emori et~al.}(2013)\citenamefont{Emori, Bauer, Ahn,
  Martinez, and Beach}}]{Emori2013}
\bibinfo{author}{\bibfnamefont{S.}~\bibnamefont{Emori}},
  \bibinfo{author}{\bibfnamefont{U.}~\bibnamefont{Bauer}},
  \bibinfo{author}{\bibfnamefont{S.}~\bibnamefont{Ahn}},
  \bibinfo{author}{\bibfnamefont{E.}~\bibnamefont{Martinez}}, \bibnamefont{and}
  \bibinfo{author}{\bibfnamefont{G.}~\bibnamefont{Beach}},
  \bibinfo{journal}{Nat. Mater.} \textbf{\bibinfo{volume}{12}},
  \bibinfo{pages}{611} (\bibinfo{year}{2013}).

\bibitem[{\citenamefont{Ryu et~al.}(2013)\citenamefont{Ryu, Thomas, Yang, and
  Parkin}}]{Ryu2013}
\bibinfo{author}{\bibfnamefont{K.}~\bibnamefont{Ryu}},
  \bibinfo{author}{\bibfnamefont{L.}~\bibnamefont{Thomas}},
  \bibinfo{author}{\bibfnamefont{S.}~\bibnamefont{Yang}}, \bibnamefont{and}
  \bibinfo{author}{\bibfnamefont{S.}~\bibnamefont{Parkin}},
  \bibinfo{journal}{Nat. Nanotechnol.} \textbf{\bibinfo{volume}{8}},
  \bibinfo{pages}{527} (\bibinfo{year}{2013}).

\bibitem[{\citenamefont{Heinze et~al.}(2011)\citenamefont{Heinze, von Bergmann,
  Menzel, Brede, Kubetzka, Wiesendanger, Bihlmayer, and
  Bl{\"{u}}gel}}]{Heinze2011}
\bibinfo{author}{\bibfnamefont{S.}~\bibnamefont{Heinze}},
  \bibinfo{author}{\bibfnamefont{K.}~\bibnamefont{von Bergmann}},
  \bibinfo{author}{\bibfnamefont{M.}~\bibnamefont{Menzel}},
  \bibinfo{author}{\bibfnamefont{J.}~\bibnamefont{Brede}},
  \bibinfo{author}{\bibfnamefont{A.}~\bibnamefont{Kubetzka}},
  \bibinfo{author}{\bibfnamefont{R.}~\bibnamefont{Wiesendanger}},
  \bibinfo{author}{\bibfnamefont{G.}~\bibnamefont{Bihlmayer}},
  \bibnamefont{and}
  \bibinfo{author}{\bibfnamefont{S.}~\bibnamefont{Bl{\"{u}}gel}},
  \bibinfo{journal}{Nat. Phys.} \textbf{\bibinfo{volume}{7}},
  \bibinfo{pages}{713} (\bibinfo{year}{2011}).

\bibitem[{\citenamefont{Romming et~al.}(2013)\citenamefont{Romming, Hanneken,
  Menzel, Bickel, Wolter, von Bergmann, Kubetzka, and
  Wiesendanger}}]{Romming2013}
\bibinfo{author}{\bibfnamefont{N.}~\bibnamefont{Romming}},
  \bibinfo{author}{\bibfnamefont{C.}~\bibnamefont{Hanneken}},
  \bibinfo{author}{\bibfnamefont{M.}~\bibnamefont{Menzel}},
  \bibinfo{author}{\bibfnamefont{J.~E.} \bibnamefont{Bickel}},
  \bibinfo{author}{\bibfnamefont{B.}~\bibnamefont{Wolter}},
  \bibinfo{author}{\bibfnamefont{K.}~\bibnamefont{von Bergmann}},
  \bibinfo{author}{\bibfnamefont{A.}~\bibnamefont{Kubetzka}}, \bibnamefont{and}
  \bibinfo{author}{\bibfnamefont{R.}~\bibnamefont{Wiesendanger}},
  \bibinfo{journal}{Science} \textbf{\bibinfo{volume}{341}},
  \bibinfo{pages}{636} (\bibinfo{year}{2013}).

\bibitem[{\citenamefont{Romming et~al.}(2015)\citenamefont{Romming, Kubetzka,
  Hanneken, {von Bergmann}, and Wiesendanger}}]{Romming2015}
\bibinfo{author}{\bibfnamefont{N.}~\bibnamefont{Romming}},
  \bibinfo{author}{\bibfnamefont{A.}~\bibnamefont{Kubetzka}},
  \bibinfo{author}{\bibfnamefont{C.}~\bibnamefont{Hanneken}},
  \bibinfo{author}{\bibfnamefont{K.}~\bibnamefont{{von Bergmann}}},
  \bibnamefont{and}
  \bibinfo{author}{\bibfnamefont{R.}~\bibnamefont{Wiesendanger}},
  \bibinfo{journal}{Phys. Rev. Lett.} \textbf{\bibinfo{volume}{114}},
  \bibinfo{pages}{177203} (\bibinfo{year}{2015}).

\bibitem[{\citenamefont{Chen et~al.}(2015)\citenamefont{Chen, Mascaraque,
  N'Diaye, and Schmid}}]{Chen2015}
\bibinfo{author}{\bibfnamefont{G.}~\bibnamefont{Chen}},
  \bibinfo{author}{\bibfnamefont{A.}~\bibnamefont{Mascaraque}},
  \bibinfo{author}{\bibfnamefont{A.~T.} \bibnamefont{N'Diaye}},
  \bibnamefont{and} \bibinfo{author}{\bibfnamefont{A.~K.}
  \bibnamefont{Schmid}}, \bibinfo{journal}{Appl. Phys. Lett.}
  \textbf{\bibinfo{volume}{106}}, \bibinfo{pages}{242404}
  (\bibinfo{year}{2015}).

\bibitem[{\citenamefont{Jiang et~al.}(2015)\citenamefont{Jiang, Upadhyaya,
  Zhang, Yu, Jungfleisch, Fradin, Pearson, Tserkovnyak, Wang, Heinonen
  et~al.}}]{Jiang2015}
\bibinfo{author}{\bibfnamefont{W.}~\bibnamefont{Jiang}},
  \bibinfo{author}{\bibfnamefont{P.}~\bibnamefont{Upadhyaya}},
  \bibinfo{author}{\bibfnamefont{W.}~\bibnamefont{Zhang}},
  \bibinfo{author}{\bibfnamefont{G.}~\bibnamefont{Yu}},
  \bibinfo{author}{\bibfnamefont{M.~B.} \bibnamefont{Jungfleisch}},
  \bibinfo{author}{\bibfnamefont{F.~Y.} \bibnamefont{Fradin}},
  \bibinfo{author}{\bibfnamefont{J.~E.} \bibnamefont{Pearson}},
  \bibinfo{author}{\bibfnamefont{Y.}~\bibnamefont{Tserkovnyak}},
  \bibinfo{author}{\bibfnamefont{K.~L.} \bibnamefont{Wang}},
  \bibinfo{author}{\bibfnamefont{O.}~\bibnamefont{Heinonen}},
  \bibnamefont{et~al.}, \bibinfo{journal}{Science}
  \textbf{\bibinfo{volume}{349}}, \bibinfo{pages}{283} (\bibinfo{year}{2015}).

\bibitem[{\citenamefont{Hoffmann et~al.}(2015)\citenamefont{Hoffmann,
  Weischenberg, Dup\'e, Freimuth, Ferriani, Mokrousov, and
  Heinze}}]{Hoffmann2015}
\bibinfo{author}{\bibfnamefont{M.}~\bibnamefont{Hoffmann}},
  \bibinfo{author}{\bibfnamefont{J.}~\bibnamefont{Weischenberg}},
  \bibinfo{author}{\bibfnamefont{B.}~\bibnamefont{Dup\'e}},
  \bibinfo{author}{\bibfnamefont{F.}~\bibnamefont{Freimuth}},
  \bibinfo{author}{\bibfnamefont{P.}~\bibnamefont{Ferriani}},
  \bibinfo{author}{\bibfnamefont{Y.}~\bibnamefont{Mokrousov}},
  \bibnamefont{and} \bibinfo{author}{\bibfnamefont{S.}~\bibnamefont{Heinze}},
  \bibinfo{journal}{Phys. Rev. B} \textbf{\bibinfo{volume}{92}},
  \bibinfo{pages}{020401(R)} (\bibinfo{year}{2015}).

\bibitem[{\citenamefont{Boulle et~al.}(2016)\citenamefont{Boulle, Vogel, Yang,
  Pizzini, {de Souza Chaves}, Locatelli, MenteÅŸ, Sala, Buda-Prejbeanu, Klein
  et~al.}}]{Boulle2016}
\bibinfo{author}{\bibfnamefont{O.}~\bibnamefont{Boulle}},
  \bibinfo{author}{\bibfnamefont{J.}~\bibnamefont{Vogel}},
  \bibinfo{author}{\bibfnamefont{H.}~\bibnamefont{Yang}},
  \bibinfo{author}{\bibfnamefont{S.}~\bibnamefont{Pizzini}},
  \bibinfo{author}{\bibfnamefont{D.}~\bibnamefont{{de Souza Chaves}}},
  \bibinfo{author}{\bibfnamefont{A.}~\bibnamefont{Locatelli}},
  \bibinfo{author}{\bibfnamefont{T.~O.} \bibnamefont{MenteÅŸ}},
  \bibinfo{author}{\bibfnamefont{A.}~\bibnamefont{Sala}},
  \bibinfo{author}{\bibfnamefont{L.~D.} \bibnamefont{Buda-Prejbeanu}},
  \bibinfo{author}{\bibfnamefont{O.}~\bibnamefont{Klein}},
  \bibnamefont{et~al.}, \bibinfo{journal}{Nat. Nanotechnol.}
  \textbf{\bibinfo{volume}{11}}, \bibinfo{pages}{449} (\bibinfo{year}{2016}).

\bibitem[{\citenamefont{Moreau-Luchaire
  et~al.}(2016)\citenamefont{Moreau-Luchaire, Moutafis, Reyren, Sampaio, Vaz,
  {Van Horne}, Bouzehouane, Garcia, Deranlot, Warnicke
  et~al.}}]{Moreau-Luchaire2016}
\bibinfo{author}{\bibfnamefont{C.}~\bibnamefont{Moreau-Luchaire}},
  \bibinfo{author}{\bibfnamefont{C.}~\bibnamefont{Moutafis}},
  \bibinfo{author}{\bibfnamefont{N.}~\bibnamefont{Reyren}},
  \bibinfo{author}{\bibfnamefont{J.}~\bibnamefont{Sampaio}},
  \bibinfo{author}{\bibfnamefont{C.~A.~F.} \bibnamefont{Vaz}},
  \bibinfo{author}{\bibfnamefont{N.}~\bibnamefont{{Van Horne}}},
  \bibinfo{author}{\bibfnamefont{K.}~\bibnamefont{Bouzehouane}},
  \bibinfo{author}{\bibfnamefont{K.}~\bibnamefont{Garcia}},
  \bibinfo{author}{\bibfnamefont{C.}~\bibnamefont{Deranlot}},
  \bibinfo{author}{\bibfnamefont{P.}~\bibnamefont{Warnicke}},
  \bibnamefont{et~al.}, \bibinfo{journal}{Nat. Nanotechnol.}
  \textbf{\bibinfo{volume}{11}}, \bibinfo{pages}{444} (\bibinfo{year}{2016}).

\bibitem[{\citenamefont{Woo et~al.}(2016)\citenamefont{Woo, Litzius,
  Kr{\"{u}}ger, Im, Caretta, Richter, Mann, Krone, Reeve, Weigand
  et~al.}}]{Woo2016}
\bibinfo{author}{\bibfnamefont{S.}~\bibnamefont{Woo}},
  \bibinfo{author}{\bibfnamefont{K.}~\bibnamefont{Litzius}},
  \bibinfo{author}{\bibfnamefont{B.}~\bibnamefont{Kr{\"{u}}ger}},
  \bibinfo{author}{\bibfnamefont{M.-Y.} \bibnamefont{Im}},
  \bibinfo{author}{\bibfnamefont{L.}~\bibnamefont{Caretta}},
  \bibinfo{author}{\bibfnamefont{K.}~\bibnamefont{Richter}},
  \bibinfo{author}{\bibfnamefont{M.}~\bibnamefont{Mann}},
  \bibinfo{author}{\bibfnamefont{A.}~\bibnamefont{Krone}},
  \bibinfo{author}{\bibfnamefont{R.~M.} \bibnamefont{Reeve}},
  \bibinfo{author}{\bibfnamefont{M.}~\bibnamefont{Weigand}},
  \bibnamefont{et~al.}, \bibinfo{journal}{Nat. Mater.}
  \textbf{\bibinfo{volume}{15}}, \bibinfo{pages}{501} (\bibinfo{year}{2016}).

\bibitem[{\citenamefont{Kurz et~al.}(2001)\citenamefont{Kurz, Bihlmayer, Hirai,
  and Bl\"ugel}}]{Kurz2001}
\bibinfo{author}{\bibfnamefont{P.}~\bibnamefont{Kurz}},
  \bibinfo{author}{\bibfnamefont{G.}~\bibnamefont{Bihlmayer}},
  \bibinfo{author}{\bibfnamefont{K.}~\bibnamefont{Hirai}}, \bibnamefont{and}
  \bibinfo{author}{\bibfnamefont{S.}~\bibnamefont{Bl\"ugel}},
  \bibinfo{journal}{Phys. Rev. Lett.} \textbf{\bibinfo{volume}{86}},
  \bibinfo{pages}{1106} (\bibinfo{year}{2001}).

\bibitem[{\citenamefont{Yoshida et~al.}(2012)\citenamefont{Yoshida, Schr\"oder,
  Ferriani, Serrate, Kubetzka, von Bergmann, Heinze, and
  Wiesendanger}}]{PhysRevLett.108.087205}
\bibinfo{author}{\bibfnamefont{Y.}~\bibnamefont{Yoshida}},
  \bibinfo{author}{\bibfnamefont{S.}~\bibnamefont{Schr\"oder}},
  \bibinfo{author}{\bibfnamefont{P.}~\bibnamefont{Ferriani}},
  \bibinfo{author}{\bibfnamefont{D.}~\bibnamefont{Serrate}},
  \bibinfo{author}{\bibfnamefont{A.}~\bibnamefont{Kubetzka}},
  \bibinfo{author}{\bibfnamefont{K.}~\bibnamefont{von Bergmann}},
  \bibinfo{author}{\bibfnamefont{S.}~\bibnamefont{Heinze}}, \bibnamefont{and}
  \bibinfo{author}{\bibfnamefont{R.}~\bibnamefont{Wiesendanger}},
  \bibinfo{journal}{Phys. Rev. Lett.} \textbf{\bibinfo{volume}{108}},
  \bibinfo{pages}{087205} (\bibinfo{year}{2012}).

\bibitem[{\citenamefont{Hardrat et~al.}(2009)\citenamefont{Hardrat, Al-Zubi,
  Ferriani, Bl\"ugel, Bihlmayer, and Heinze}}]{Hardrat:09.1}
\bibinfo{author}{\bibfnamefont{B.}~\bibnamefont{Hardrat}},
  \bibinfo{author}{\bibfnamefont{A.}~\bibnamefont{Al-Zubi}},
  \bibinfo{author}{\bibfnamefont{P.}~\bibnamefont{Ferriani}},
  \bibinfo{author}{\bibfnamefont{S.}~\bibnamefont{Bl\"ugel}},
  \bibinfo{author}{\bibfnamefont{G.}~\bibnamefont{Bihlmayer}},
  \bibnamefont{and} \bibinfo{author}{\bibfnamefont{S.}~\bibnamefont{Heinze}},
  \bibinfo{journal}{Phys. Rev. B} \textbf{\bibinfo{volume}{79}},
  \bibinfo{pages}{094411} (\bibinfo{year}{2009}).

\bibitem[{\citenamefont{Al-Zubi et~al.}(2011)\citenamefont{Al-Zubi, Bihlmayer,
  and Bl\"ugel}}]{Al-Zubi:11.2}
\bibinfo{author}{\bibfnamefont{A.}~\bibnamefont{Al-Zubi}},
  \bibinfo{author}{\bibfnamefont{G.}~\bibnamefont{Bihlmayer}},
  \bibnamefont{and} \bibinfo{author}{\bibfnamefont{S.}~\bibnamefont{Bl\"ugel}},
  \bibinfo{journal}{Phys. Status Solidi B} \textbf{\bibinfo{volume}{248}},
  \bibinfo{pages}{2242} (\bibinfo{year}{2011}).

\bibitem[{SI()}]{SI}
\bibinfo{howpublished}{Supplemental material.}

\bibitem[{\citenamefont{Bode}(2003)}]{Bode2003a}
\bibinfo{author}{\bibfnamefont{M.}~\bibnamefont{Bode}}, \bibinfo{journal}{Rep.
  Prog. Phys.} \textbf{\bibinfo{volume}{66}}, \bibinfo{pages}{523}
  (\bibinfo{year}{2003}).

\bibitem[{\citenamefont{Wiesendanger}(2009)}]{Wiesendanger2009}
\bibinfo{author}{\bibfnamefont{R.}~\bibnamefont{Wiesendanger}},
  \bibinfo{journal}{Rev. Mod. Phys.} \textbf{\bibinfo{volume}{81}},
  \bibinfo{pages}{1495} (\bibinfo{year}{2009}).

\bibitem[{spi()}]{spindynamics}
\bibinfo{howpublished}{With an external magnetic field a skyrmion lattice phase
  can in principle be induced as shown by spin dynamics simulations (see
  \cite{SI}), however, due to the deep energy minimum the transition fields are
  about 50~T.}

\bibitem[{\citenamefont{Bode et~al.}(2002)\citenamefont{Bode, Heinze, Kubetzka,
  Pietzsch, Nie, Bihlmayer, Bl{\"{u}}gel, and Wiesendanger}}]{Bode2002}
\bibinfo{author}{\bibfnamefont{M.}~\bibnamefont{Bode}},
  \bibinfo{author}{\bibfnamefont{S.}~\bibnamefont{Heinze}},
  \bibinfo{author}{\bibfnamefont{A.}~\bibnamefont{Kubetzka}},
  \bibinfo{author}{\bibfnamefont{O.}~\bibnamefont{Pietzsch}},
  \bibinfo{author}{\bibfnamefont{X.}~\bibnamefont{Nie}},
  \bibinfo{author}{\bibfnamefont{G.}~\bibnamefont{Bihlmayer}},
  \bibinfo{author}{\bibfnamefont{S.}~\bibnamefont{Bl{\"{u}}gel}},
  \bibnamefont{and}
  \bibinfo{author}{\bibfnamefont{R.}~\bibnamefont{Wiesendanger}},
  \bibinfo{journal}{Phys. Rev. Lett.} \textbf{\bibinfo{volume}{89}},
  \bibinfo{pages}{237205} (\bibinfo{year}{2002}).

\bibitem[{\citenamefont{Hanneken et~al.}(2015)\citenamefont{Hanneken, Otte,
  Kubetzka, Dup\'e, Romming, von Bergmann, Wiesendanger, and
  Heinze}}]{Hanneken:15.1}
\bibinfo{author}{\bibfnamefont{C.}~\bibnamefont{Hanneken}},
  \bibinfo{author}{\bibfnamefont{F.}~\bibnamefont{Otte}},
  \bibinfo{author}{\bibfnamefont{A.}~\bibnamefont{Kubetzka}},
  \bibinfo{author}{\bibfnamefont{B.}~\bibnamefont{Dup\'e}},
  \bibinfo{author}{\bibfnamefont{N.}~\bibnamefont{Romming}},
  \bibinfo{author}{\bibfnamefont{K.}~\bibnamefont{von Bergmann}},
  \bibinfo{author}{\bibfnamefont{R.}~\bibnamefont{Wiesendanger}},
  \bibnamefont{and} \bibinfo{author}{\bibfnamefont{S.}~\bibnamefont{Heinze}},
  \bibinfo{journal}{Nature Nanotech.} \textbf{\bibinfo{volume}{10}},
  \bibinfo{pages}{1039} (\bibinfo{year}{2015}).

\bibitem[{\citenamefont{Tersoff and Hamann}(1985)}]{TH-model}
\bibinfo{author}{\bibfnamefont{J.}~\bibnamefont{Tersoff}} \bibnamefont{and}
  \bibinfo{author}{\bibfnamefont{D.~R.} \bibnamefont{Hamann}},
  \bibinfo{journal}{Phys. Rev. B} \textbf{\bibinfo{volume}{31}},
  \bibinfo{pages}{805} (\bibinfo{year}{1985}).

\bibitem[{\citenamefont{Wortmann et~al.}(2001)\citenamefont{Wortmann, Heinze,
  Kurz, Bihlmayer, and Bl\"ugel}}]{PhysRevLett.86.4132}
\bibinfo{author}{\bibfnamefont{D.}~\bibnamefont{Wortmann}},
  \bibinfo{author}{\bibfnamefont{S.}~\bibnamefont{Heinze}},
  \bibinfo{author}{\bibfnamefont{P.}~\bibnamefont{Kurz}},
  \bibinfo{author}{\bibfnamefont{G.}~\bibnamefont{Bihlmayer}},
  \bibnamefont{and} \bibinfo{author}{\bibfnamefont{S.}~\bibnamefont{Bl\"ugel}},
  \bibinfo{journal}{Phys. Rev. Lett.} \textbf{\bibinfo{volume}{86}},
  \bibinfo{pages}{4132} (\bibinfo{year}{2001}).

\bibitem[{\citenamefont{von Bergmann et~al.}(2012)\citenamefont{von Bergmann,
  Menzel, Serrate, Yoshida, Schr\"oder, Ferriani, Kubetzka, Wiesendanger, and
  Heinze}}]{Bergmann2012}
\bibinfo{author}{\bibfnamefont{K.}~\bibnamefont{von Bergmann}},
  \bibinfo{author}{\bibfnamefont{M.}~\bibnamefont{Menzel}},
  \bibinfo{author}{\bibfnamefont{D.}~\bibnamefont{Serrate}},
  \bibinfo{author}{\bibfnamefont{Y.}~\bibnamefont{Yoshida}},
  \bibinfo{author}{\bibfnamefont{S.}~\bibnamefont{Schr\"oder}},
  \bibinfo{author}{\bibfnamefont{P.}~\bibnamefont{Ferriani}},
  \bibinfo{author}{\bibfnamefont{A.}~\bibnamefont{Kubetzka}},
  \bibinfo{author}{\bibfnamefont{R.}~\bibnamefont{Wiesendanger}},
  \bibnamefont{and} \bibinfo{author}{\bibfnamefont{S.}~\bibnamefont{Heinze}},
  \bibinfo{journal}{Phys. Rev. B} \textbf{\bibinfo{volume}{86}},
  \bibinfo{pages}{134422} (\bibinfo{year}{2012}).

\bibitem[{\citenamefont{Kurz et~al.}(2004)\citenamefont{Kurz, F\"orster,
  Nordstr\"om, Bihlmayer, and Bl\"ugel}}]{Kurz2004}
\bibinfo{author}{\bibfnamefont{P.}~\bibnamefont{Kurz}},
  \bibinfo{author}{\bibfnamefont{F.}~\bibnamefont{F\"orster}},
  \bibinfo{author}{\bibfnamefont{L.}~\bibnamefont{Nordstr\"om}},
  \bibinfo{author}{\bibfnamefont{G.}~\bibnamefont{Bihlmayer}},
  \bibnamefont{and} \bibinfo{author}{\bibfnamefont{S.}~\bibnamefont{Bl\"ugel}},
  \bibinfo{journal}{Phys. Rev. B} \textbf{\bibinfo{volume}{69}},
  \bibinfo{pages}{024415} (\bibinfo{year}{2004}).

\end{thebibliography}
\end{document}